\documentclass[pdflatex,sn-mathphys-num]{sn-jnl}

\usepackage{graphicx}
\usepackage{multirow}
\usepackage{amsmath,amssymb,amsfonts}
\usepackage{amsthm}
\usepackage{mathrsfs}
\usepackage[title]{appendix}
\usepackage{xcolor}
\usepackage{textcomp}
\usepackage{manyfoot}
\usepackage{booktabs}
\usepackage{algorithm}
\usepackage{algorithmicx}
\usepackage{algpseudocode}
\usepackage{listings}
\usepackage{xcolor}
\usepackage{makecell}

\theoremstyle{thmstyleone}%
\theoremstyle{thmstyletwo}%

\theoremstyle{thmstylethree}%

\begin{document}

\title[Article Title]{Operationalizing Governed Agentic Workflows in Regulated Finance: A Customer Zero Case Study using Zafin AIOS}


\author*[1]{\fnm{Shahir} \sur{Daya}}\email{shahir.daya@zafin.com}
\author*[2]{\fnm{Rob} \sur{Dunlap}}\email{rob.dunlap@zafin.com}
\author*[3]{\fnm{Branavan} \sur{Selvasingham}}\email{branavan.selvasingham@zafin.com}
\author*[4]{\fnm{Alexandre} \sur{Boulenger}}\email{alexandre.boulenger@zafin.com}
\author*[5]{\fnm{Ahmed} \sur{Abdelmoneim}}\email{ahmed.abdelmoneim@zafin.com}
\author*[6]{\fnm{Salvatore} \sur{Vella}}\email{sal.vella@zafin.com}



\affil*[1]{\orgdiv{Chief Technology Officer}, \orgname{Zafin}, \orgaddress{\city{Toronto}, \state{Ontario}, \country{Canada}}}
\affil*[2]{\orgdiv{Chief Product Officer}, \orgname{Zafin}, \orgaddress{\city{Toronto}, \state{Ontario}, \country{Canada}}}
\affil*[3]{\orgdiv{VP, AI and Product Engineering}, \orgname{Zafin}, \orgaddress{\city{Toronto}, \state{Ontario}, \country{Canada}}}
\affil*[4]{\orgdiv{Head of AIOS Product Innovation}, \orgname{Zafin}, \orgaddress{\city{Dubai}, \state{Dubai}, \country{United Arab Emirates (UAE)}}}
\affil*[5]{\orgdiv{AI Researcher}, \orgname{Zafin}, \orgaddress{\city{Toronto}, \state{Ontario}, \country{Canada}}}
\affil*[6]{\orgdiv{AI Researcher}, \orgname{Zafin}, \orgaddress{\city{Toronto}, \state{Ontario}, \country{Canada}}}





\abstract{

Regulated industries such as finance and healthcare face critical governance challenges that hamper the adoption of AI-assisted software development and hinder the transition from experimental pilots to production work. Enterprises need to leverage the power and speed of AI-assisted development while maintaining compliance, traceability, and adherence to AI standards and frameworks. This paper presents Zafin AIOS, an end-to-end orchestration platform and control plane designed to deploy AI within regulated industries such as finance. We provide the architectural principles and an empirical comparison between AI-assisted and human-led workloads. To validate this framework empirically, we conduct a case study analysis of Zafin's internal enterprise workflows, which we refer to as Customer Zero. This paper contributes a production case study of AI-assisted software engineering in regulated industries. We evaluate the results of using Zafin AIOS with two different production codebases, one for the pricing execution platform and the other for the client deal negotiation platform. Our contribution is a customer case study for regulated industries with production codebases and governance, which includes proof of work. The results demonstrate that strictly governed agentic orchestration increases enterprise capacity without compromising control. In the pricing execution platform workflows, the platform provided complete observability and reduced effort by 38\%. In the client deal negotiation platform, it reduced reporting work by 76\% while maintaining all predefined data privacy, role-based access, and audit guardrails. AIOS makes agentic work observable and auditable through governed orchestration, proof of work, and human reviews. This paper bridges the gap between AI-assisted software engineering theory and the operational reality of compliance. Closing this gap is important to help enterprises take advantage of AI technologies. These results are descriptive of the benefits of using governed workflows in regulated industries and do not yet establish causal productivity gains against a human-led baseline.

}




\maketitle

\section{Introduction}\label{sec1}

The introduction of Generative AI into the software development process has generated great interest and promises to redefine how and how quickly software is developed. Code models and systems such as OpenAI's Codex, GitHub Copilot, Claude Code, and Antigravity have shown a remarkable ability to translate natural-language intent into running code, whether for fixing defects or adding new features. Beyond dramatic speed and throughput gains, these systems lower the barrier to entry for non-technical users.

Many enterprises are embarking on pilots. Moving these pilots into large-scale production, especially for regulated industries, remains severely hampered. One problem is that these industries require deterministic behavior, whereas Generative AI is inherently probabilistic. In addition, these enterprises require auditability, data privacy, and adherence to security standards. The codebases these tools run on are valuable assets, often created and maintained for decades, with values in the billions of dollars. These assets are important and subject to strict governance requirements. The gap to address is how to balance the speed and agility Generative AI offers with the control and governance enterprises require.

The proposed solution is AIOS, an end-to-end orchestration platform and control plane designed to enable and scale Generative AI in regulated industries.

AIOS is based on four core architectural principles that guide its design and development.

\begin{enumerate}
    \item \textbf{Governed Orchestration}. Intelligent routing of tasks across models and agents based on the risk and complexity of the task to be performed.
    \item \textbf{Proof of Work}. All AI actions have deterministic traceability and auditability for later analysis and review.
    \item \textbf{Human Authority}. Humans remain paramount to the solution, and human-in-the-loop reviews are available for high-risk and high-stakes changes.
    \item \textbf{Granular Cost Controls}. With concerns about inference costs, the system lets you specify budgeting trade-offs to ensure the right models and costs are used for the task.
\end{enumerate}

\begin{figure}[t]
  \centering
  \includegraphics[width=\linewidth]{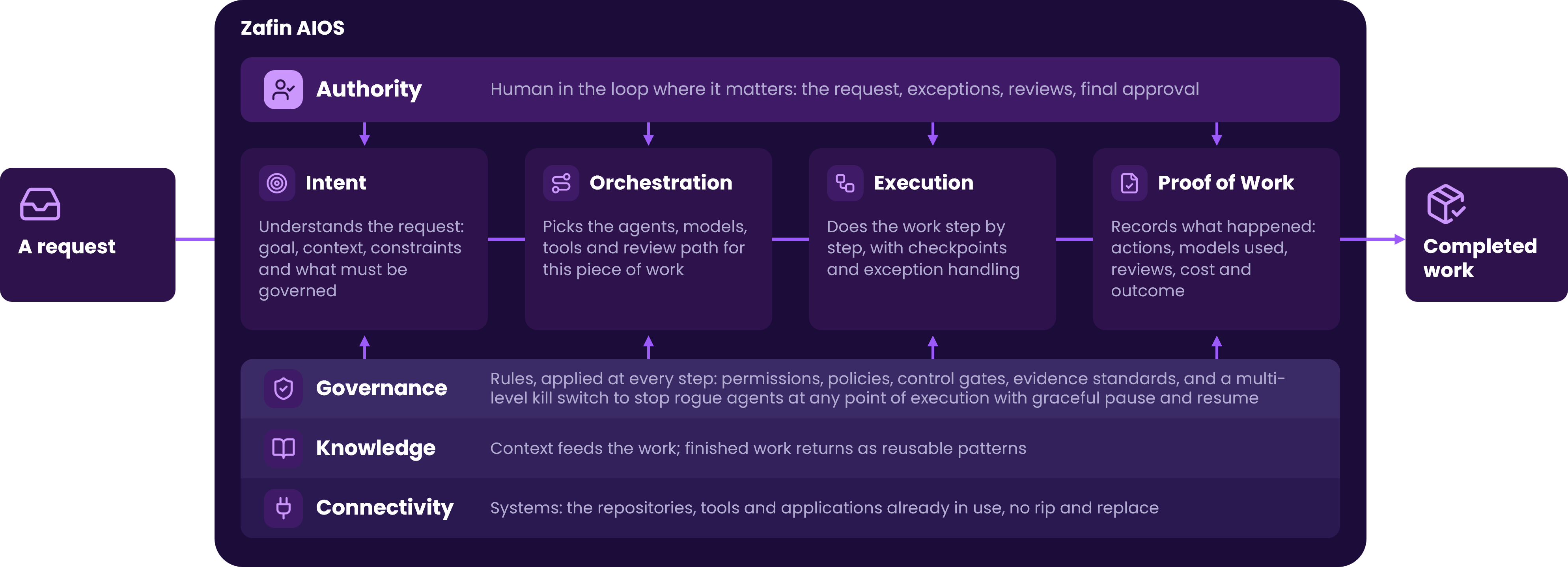}
  \caption{Zafin AIOS High-Level Architecture}
  \label{fig:AIOS-Architecture}
\vspace{-12pt}
\end{figure}

Figure \ref{fig:AIOS-Architecture} shows the high-level architecture of Zafin AIOS.

The novelty of this work, unlike other studies that use theoretical, sandbox, or small codebases, is its study of Generative AI in high-value production codebases, with practitioner estimates of the percentage of total effort saved by AIOS during processing. This allows comparison of the benefits of Generative AI and the issues that must be resolved for the technology to be adopted in regulated industries. To our knowledge, this is a unique study, as we have not found others in the literature.

This paper studies and presents two production workloads.

\begin{enumerate}
    \item The pricing execution platform is an application focused on complex pricing execution involving many product domains across a multi-million line of code codebases. For this codebase, we are exploring the use of AIOS to perform software engineering tasks focused on adding new features. The programming language for this codebase is Java.
    \item The client deal negotiation platform is a predominantly front-end application that provides complex banker experiences and integrates with many other apps. The programming language for this codebase is TypeScript.
\end{enumerate}

These two codebases allow us to study the impact of AI-assisted software development for both a complex backend codebase and a less complex, integrated client deal negotiation platform.

The core argument that governs orchestration is that it can meaningfully increase the capacity to use Generative AI tools without compromising control and governance.

For the pricing execution platform codebase, effort is estimated to drop by 38\%. For the client deal negotiation platform, effort decreased by 76\%. In both cases, we achieved this while fully maintaining privacy, role-based access controls, and complete auditability.

The contributions of the paper are twofold:

\begin{enumerate}
    \item \textbf{Introduction of Zafin AIOS}, an end-to-end orchestration platform for software engineering with governance and controls required for regulated industries such as banking.
    \item \textbf{Empirical Evaluation of AIOS} compares the effort for several features implemented, using estimates from the development team and their supervisors to reach consensus.
\end{enumerate}

The rest of the paper follows. Section \ref{sec-literature-review} provides a summary of the Literature Review of related work, Section \ref{sec-aios-architecture} describes the Zafin AIOS system used in this study, Section \ref{sec-methodology} provides the Methodology used for the Customer Zero study, Section \ref{sec-results} provides the Results, Section \ref{sec-discussion} provides a discussion of the results, and Section \ref{sec-conclusion} provides a conclusion and positioning of the importance of the work.

\section{Literature Review}\label{sec-literature-review}

\subsection{The Tension of Artificial Intelligence in Software Development}

The integration of AI-assisted tools into enterprise software development is generating unprecedented enthusiasm. This revolution could dramatically improve software development by increasing speed and quality. A dramatic step-function increase in developer productivity \cite{zheng_survey_2023} \cite{lo_trustworthy_2023} can assist with bug fixes, new feature development, and documentation generation \cite{zheng_survey_2023}. Empirical evidence also shows that these tools can lower barriers to entry, allowing non-technical members to code \cite{pimenova_good_2025} \cite{gadde_democratizing_2025}. Developers enter a flow state where rapid prototyping and quick implementation of alternatives reduce timelines from weeks to hours \cite{lo_trustworthy_2023} \cite{pimenova_good_2025}.

A gap remains between AI-assisted tools used in sandbox pilots and product-ready environments in finance and healthcare \cite{haque_llms_2025}. In these environments, the cost of failure is very high, where code bugs or security vulnerabilities can lead to large regulatory penalties and corporate reputation loss \cite{alebachew_ai-guided_2025} \cite{haque_sok_2025} \cite{anderson_beyond_2025}. Whereas non-regulated industries can move fast, enterprises in regulated industries often get stuck in a pilot trap, unable to move to production because these tools lack the necessary quality, compliance, and security tooling \cite{pandita_towards_2018} \cite{sharma_beyond_2025}.

This governance gap needs to be addressed. This is a tension between the probabilistic nature of AI-assisted development and the more strictly deterministic requirements of regulatory compliance \cite{bandara_agentsway_2025} \cite{xing_when_2025}. Leaving process enforcement to a stochastic LLM is unacceptable in regulated environments. Enterprises in regulated industries urgently need an explicit, executable control plane to bridge this governance gap.

\subsection{Current State of AI-Assisted Software Development}

The current state of AI-assisted development is transforming the software engineer's role from writing individual lines of code to defining intent and mediating interaction with AI \cite{karpathy_tweet_2025} \cite{wang_vibecontract_2026} \cite{meske_vibe_2025} \cite{treude_how_2025}.

The model context protocol (MCP) is an important evolution that standardizes interfaces between the LLM and enterprise tools \cite{gaire_systematization_2025}. From a governance perspective, this layer provides a central governance enforcement layer for access to enterprise resources.

\subsection{Governance Challenges in High-Value Production Codebases}

In high-value codebases, codebase integrity is critical. Human review of all of the generated code is difficult.

There are issues of software vulnerabilities introduced in AI-assisted code. These have been classified of (a) broken access, where AI can generate code that bypasses authentication layers, (b) package source leaks, where incorrect or insecure packages are introduced into the software supply chain, and (c) prompt injection into the code generation process, where the intent may be manipulated before code is generated \cite{deng_understanding_2026} \cite{waseem_vibe_2025} \cite{zhao_is_2026}.

To mitigate these risks, research argues that agents must operate in a zero-trust environment, meaning every agent action must be verified and agents must be restricted by the principle of least privilege \cite{brandes_custody_2026}.

Another condition to mitigate is agent entropy, which can cause code quality to decay and architecture to drift over time. Without governing this entropy, the codebase can become unmaintainable. Because this entropy is a feature of AI-assisted coding, a necessary transition from vibe coding to enterprise engineering standards requires governed orchestration and more formal specification \cite{casserini_beyond_2026}.

\subsection{Architectural Pillars for Enterprise AI Acceptance}

Transforming AI-assisted development so enterprises in regulated industries can rely on it requires four architectural pillars to ground stochastic vibe coding in a more deterministic engineering discipline.

\begin{enumerate}
    \item \textbf{Governed Orchestration}. Orchestration frameworks need to incorporate hard guardrails and policy as code \cite{gaire_systematization_2025} \cite{brandes_custody_2026} \cite{zhu_specification_2026} \cite{vella2026ateam}. High-risk and high-compliance changes have stricter workflows than simple changes \cite{wang_empirical_2025} \cite{miculicich_veriguard_2025} \cite{wang_reflection-driven_2025}. Another safety mechanism is safety via debate, where an adversarial approach allows agents to find flaws in each other's reasoning \cite{irving_ai_2018}.
    \item \textbf{Human-In-The-Loop Requirements}. Despite autonomous agents' capabilities, research argues that some changes need human oversight, so any framework must include a flexible workflow that lets software engineers remain part of the development process \cite{shuvo_context_2026}.
    \item \textbf{Specification Driven Development}. This shifts the specification and other grounded artifacts to be the source of truth rather than the code \cite{mitchell_position_2025} \cite{alenezi_ai-driven_2025} \cite{wang_vibecontract_2026} \cite{vella2026axl}. AI and optional human oversight ensure the code adheres to the engineering specification. 
    \item \textbf{Telemetry and Auditing}. Regulatory compliance requires that all agentic actions be auditable. This provides a forensic trail of every decision made \cite{liu_large_2025} \cite{fatouros_organization-scoped_2026}. This enables further investigation and learning that can be incorporated into the orchestration and governance process.
\end{enumerate}

The successful integration of AI-assisted development into regulated industries depends on a robust framework that bridges the gap between the speed and flexibility of unconstrained vibe coding and the deterministic and governed needs of regulated industries.

\section{The Zafin AIOS Architecture}\label{sec-aios-architecture}

\subsection{Introduction}

Zafin AIOS provides a control plane that governs AI agents, LLM models, tools the LLMs can use, skills available to the LLMs and tools, and knowledge provided to the system and learned through interactions and experience.

Agents are modeled much like human employees with the same security controls and access to knowledge they can use. Least privlege role-base access control (RBAC) is used and every action traces back to an accountable entity.

Figure \ref{fig:AIOS-Architecture} provides a high-level picture of the AIOS architecture.

Work is routed to AIOS through structured Intents. LLM models and tools are registered in the Agent registry, and each agent decides which LLM model to pick for the required task, either deterministically or based on the task.

Governance is enforced at execution time. Agents get identities in the same corporate directory as staff, under least-privilege RBAC, and an egress proxy caps what a running agent can reach. No shared or anonymous agent principal exists, so every action traces back to an accountable identity.

\begin{figure}[h!]
  \centering
  \includegraphics[width=\linewidth]{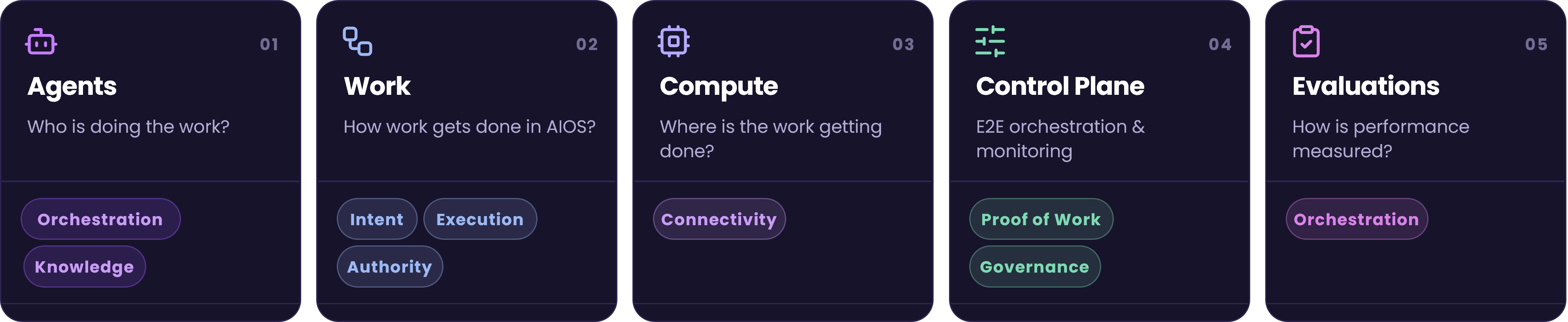}
  \caption{Elements of the AIOS System}
  \label{fig:AIOS-Element}
\vspace{-12pt}
\end{figure}

Figure \ref{fig:AIOS-Element} shows the elements of the AIOS System and the architecture levels each maps to. We will discuss each in turn.

At the center of the system are the agents. An agent is the unit that performs work. Each is assigned an identity, a role, and a scoped set of skills, tools, and knowledge, and it executes the steps of a work item under the governing workflow. Agents are modeled much like human employees operating under least-privilege role-based access control (RBAC). Every action traces back to an accountable identity. Each agent operates inside the bounds the workflow and policies define.

The work is specified with a structured input that provides a case to the system. The agent work management system provides a managed environment where work is governed through a multi-step workflow where every step is tracked to completion.

The outer workflow is configurable and deterministic, and the agent interior is not. Step order and gate placement are fixed at the workflow level, while the agent chooses its own approach within a step and must finish with whatever evidence that step demands.

A run can be replayed from the recorded state. Steps are deterministic, and transitions to the workflow are from the specified workflow and policies. These workflows are versioned so that older workflows remain reviewable and auditable.

\begin{figure}[h!]
  \centering
  \includegraphics[width=\linewidth]{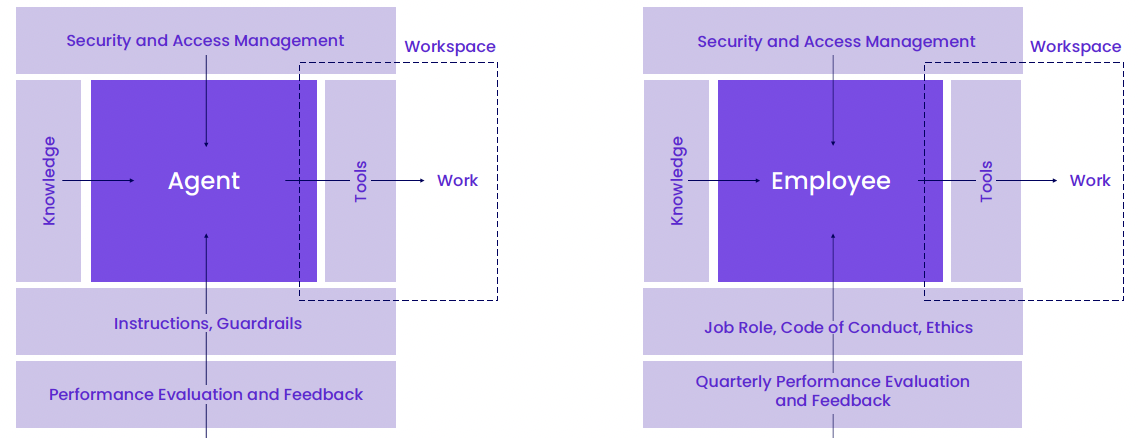}
  \caption{What is an Agent? AI Agents and Employees share significant similarities.}
  \label{fig:Cust0-Agent}
\vspace{-12pt}
\end{figure}

Figure \ref{fig:Cust0-Agent} shows the elements of an AI Agent within the AIOS system. Note the similarity between modeling AI agents and human employees: both use the same security and access controls, access to knowledge, and regular performance evaluation to tune performance.

The Knowledge Fabric supplies grounding for the system to use. These include source code repositories and data stores, including relational, vector, and graph stores. Agents pull the approved institutional context when a work item starts. Skills and tools come from their own registries, scoped to the profile.

The fabric gets written to as well as read. An agent that finishes cleanly authors a case study, an operator reviews it, and it is published back into the fabric as retrievable context for later runs. This knowledge allows the system to learn over time.

Compute usage depends on the task. Models can be explicitly chosen for the task, or the agent is allowed to choose the model - expensive or less expensive, cloud or on-premises- based on policy and task.

The Control Plan provides end-to-end orchestration and monitoring for the system. 

Every harness invocation goes through an OpenTelemetry tracer proxy that records per-turn token counts, model choice, and latency, and flags stuck sessions when heartbeats are missed. Human operators drill from work item to step to turn and read the exact prompt and the output it produced. That same stream prices tokens at configurable per-model rates and attributes spend by agent, work item, model, and day, with compute billed from the provider or amortized from hardware.

The system provides analytics to help understand success metrics. Costing metrics are also provided to understand changes in models and strategy that could be used.

Role-Based Access Control (RBAC) controls agent access, giving them the minimum access needed to do their work.

Proof of work is generated automatically: the plan and its phase rationale, a transcript per step, the acceptance verdict, a walkthrough of what changed, plus tests, screenshots, and API traces. 

Human authority shows up as gates at intake, plan, and result, each with a rework path back to the agent. An agent that hits ambiguity mid-step calls a question tool; the item parks in awaiting-human, and execution resumes once an answer lands. Gate decisions and profile changes are timestamped.

Evaluation keeps running after deployment. Benchmark sets stay fixed, rubrics weight deterministic signals against cost and LLM-as-judge criteria, and evals fire inline, ad hoc, on a drift schedule, and across models in experiments. Scheduled replay is what makes supervisory oversight defensible once per-item review stops scaling.

By cross-referencing the Zafin AIOS architecture with the architecture challenges identified in the Literature Review, we can see how AIOS transforms theoretical requirements into a running control plan.

\begin{itemize}
    \item \textbf{Specification-Driven Routing}. Work is provided through structured intents and a deterministic workflow with human approvals and gates. This ensures AI agents do not deviate from the provided specifications.
    \item \textbf{Governed Orchestration and Routing}. The AIOS control plane enforces least-privilege RBAC and limits an agent's Egress. The execution is bound to the approved boundaries.
    \item \textbf{Enforced Human Activity}. AIOS mandates (based on risk) review gates at case intake, planning, and execution. When ambiguity arises, the system parks the workflow in a waiting-for-human-input state.
    \item \textbf{Embedded Proof of Work}. Using the OpenTelemetry trace, every step of the workflow, from rationale to API calls and token exchanges with LLMs, is deterministically captured for later audit.
    \item \textbf{Granular Cost Controls}. The monitoring captures not only execution but also token usage per model, AI agent, and case, enabling inference budgeting and use of a variety of models, whether on-premises or in the cloud.
\end{itemize}

The Zafin AIOS system makes AI-assisted software development a governed engineering process for probabilistic AI agents. It meets the needs of enterprises in regulated environments by providing deterministic workflows, human reviews and gates, and proof of work and auditability. In addition, cost monitoring gives the enterprise with visibility and options with LLM models and work placement. Grounded enterprise knowledge and traceability back to an entity with strict RBAC controls allow full traceability and auditability.

\section{Methodology}\label{sec-methodology}

This case study uses a comparative customer zero design. It focuses on Zafin's own internal production codebases rather than on synthetic benchmarks. This case study is conducted within Zafin, a regulated FinTech, on its own live production codebases

We compare AIOS-assisted execution versus human estimates of how long the same work would have taken. The work is from design to production of the features.

We treat the results as descriptive productivity evidence, not a controlled human-led versus AI-assisted baseline. To our knowledge, few studies examine AI-assisted software engineering in live regulated production codebases.

\subsection{Study Setting}

Zafin is a FinTech in the financial services industry. As such, its codebases and software engineering practices must adhere to regulated-industry guidelines.

A unique aspect of this case study is the use of live, evolving production codebases rather than synthetic benchmarks or hackathons.

AIOS was also under development, with features also evolving. The features chosen were over 2 months old. Re-running the study as the AIOS codebase evolves would be another option.

This study examines productivity improvements in regulated industries with production code. This means not only can the system generate the code, but it also has the governance constraints - proof of work, auditability, human oversight - that are required for regulated industries.

\subsection{Codebases}

This study uses two production codebases.

\begin{enumerate}
    \item \textbf{pricing execution platform}. In this engine, you can configure complex rules and logic for banking products and perform the execution of policies across many complex components. This often includes many levels of decision-making across many rules and jobs to determine the exact impacts. The platform handles very large enterprise volumes (hundreds of millions of accounts) and executes within hours to ensure banking service levels are met. It is a large, complex codebase with millions of lines of code.
    \item \textbf{client deal negotiation platform}. This application allows a banker to negotiate complex pricing deals across banking domains. It is a heavily user-interface codebase that prioritizes intuitive experiences for bankers in complex processes. This is a less complex codebase but still large with millions of lines of code.
\end{enumerate}

We selected these two codebases as they allow us to study the impact of AI-assisted software development for both a complex, backend codebase and for a less complex, though large, user interface application.

In terms of programming languages, the pricing execution platform is primarily Java based with significant SQL code and the client deal negotiation platform is mainly TypeScript and user interfaces (UI).

\subsection{Case Selection}

Enhancement requests (ERs) were selected from real work items in the backlog. For the pricing execution platform, we selected 30 ERs, while for the client deal negotiation platform, we selected 19 ERs. This gave a total of 49 enhancement requests for this study.

These requests represented work that previously would have been done solely by human developers. These are real, live enhancement requests and not synthetic benchmarks. This sample provides us with real-world measurements.

These items cover multiple products and multiple codebases rather than a single isolated system. The selected cases include varied task sizes and complexity levels.

\subsection{Unit of Analysis}

We treated each Enhancement Request as a single case. The ER started with the Product Manager describing the outcome in customer terms, not engineering terms.

For each ER, we captured the product/codebase, the contribution to the overall work (from design to production), the total number of execution cycles (retries), and the number of human retries. We also measured the total elapsed time for the request.

This data allows results to be summarized both overall and by product/codebase.

\subsection{Baseline}

The baseline was based on expert developer estimates. Estimates reflect the effort required without AIOS and the effort performed by automation versus human-led. The team leader's consensus was used to double-check the developers' estimates.

The study compares the estimated effort avoided against a human-led implementation for each task. It is performed on real tasks, production codebases, and ERs of varying sizes and complexities. 

\subsection{Metrics Chosen}

The following metrics are what we would ideally have to be able to support the claims in the paper:

\begin{enumerate}
    \item Estimated effort reduction: Estimated effort reduction refers to the percentage of total work from design through production that practitioners estimated was avoided through AIOS assistance. 
    \item Execution cycles: Execution cycles refer to the number of AIOS and human workflow iterations required to complete an enhancement request.
    \item User-initiated interventions: Number of times a human had to steer, clarify, or intervene in the AI-assisted workflow.
    \item Governance preservation: Whether privacy, Role-based Access Control (RBAC), auditability, and proof-of-work controls remained intact.
\end{enumerate}

\subsection{Governance Controls}

The following Governance controls were in place for all enhancement requests processed:

\begin{enumerate}
    \item Work entered AIOS through structured intents.
    \item Agents operated under least-privilege role-based access control.
    \item Workflow defined and human-reviewable to ensure that high-risk and more complex tasks receive more reviews than simple tasks.
    \item Egress and tool access were governed.
    \item Human review gates were available at intake, planning, and result stages.
    \item Ambiguous work could pause for human clarification.
    \item Actions were logged to create proof of work.
    \item The workflow produced artifacts such as plans, transcripts, acceptance evidence, tests, screenshots, traces, and review records.
\end{enumerate}

\subsection{Analysis Approach}

Results are grouped by product/codebase. Average AIOS savings are calculated across all ERs available.

Human oversight and cycle counts are used to understand the amount of human intervention still required.

We interpret the findings not only in terms of productivity but also in terms of the ability to meet the requirements of a regulated industry.

\subsection{Limitations}

We acknowledge several limitations in this study.

First, the baseline relies on expert estimates rather than a fully controlled, human-led comparison. We believe it is important to understand real-world improvements in regulated industries, and making parallel changes in production codebases is not feasible. With expert reviews and team leader consensus, we provide descriptive metrics and a contribution of benefits beyond what most studies provide with sandboxes and hackathons.

Second, the codebases and AIOS evolved during the study period. This is also a reality in real-world case studies. We captured ERs over a small time period of 2 months. We plan a further study to update data in the future.

\section{Results}\label{sec-results}

This section reports the observed outcomes from 49 real Enhancement Requests (ERs) processed with Zafin AIOS. The results are presented descriptively rather than as a controlled experiment.

The focus is on three items. The first is the estimated effort reduction. The second is the amount of human intervention still required. The third is whether governed execution, including user-defined workflow and proof of work, was preserved.

Effort is the total work done by the development team, from design through implementation, testing, and production. This effort savings is estimated by human experts - the development team and their supervisor for consensus.

\subsection{Samples Used}

Forty-nine ERs were analyzed. 30 requests came from the pricing execution platform codebase and 19 requests came from the client deal negotiation platform codebase.

The two codebases represent two different types of production codebases. The pricing execution platform is a large, complex codebase for banking process execution. The client deal negotiation platform is a UI-heavy codebase with less complexity but more user interface rules.

Because the impact of using Zafin AIOS may differ between codebases and tasks required in those codebases, we report both overall results and results per codebase.

\subsection{Effort Reduction}

Zafin AIOS reduced estimated effort for both production workloads and every enhancement request. The impact varied by codebase.

For the pricing execution platform, the average estimated effort reduction was 38\%. For the client deal negotiation platform, the average estimated reduction was about 76\%. We note that these results may be due to the complexity of each codebase with the pricing execution platform as more complex code and the client deal negotiation platform more UI intensive and less complex.

Across all 49 requests, the average estimated effort reduction was 53\%. These are significant effort reductions for enterprises in regulated industries.

\subsection{Human Intervention}

Human intervention occurs when a human must steer, correct, or guide the workflow. At least one human intervention means the workflow is not yet fully autonomous.

The overall average was 2.9 human interventions per request with a slight variation between codebases. The pricing execution platform average was 3 interventions per request versus the client deal negotiation platform average of 2.6 interventions per request. The median overall was 2 interventions per request.

Human intervention remained part of the workflow. These results indicate that humans were not eliminated from the workflow; instead, the role shifted to steering, clarification, and review.

\subsection{Execution Cycles}

Execution cycles are the total number of AIOS plus human workflow cycles required to complete the work. Higher cycle count may indicate more complex or ambiguous tasks. Lower cycle counts may indicate simpler tasks or tasks that can be better specified.

The overall average was 7.8 cycles per request, with a median of 5 cycles per request. Codebases differed: the pricing execution platform averaged 6.7 cycles per request, and the client deal negotiation platform had 9.4 cycles per request.

Interestingly, the client deal negotiation platform had both the higher cycle counts and the higher mean estimated effort reduction. This suggests that cycle time alone should not be interpreted as inefficiency. In some workflows, AIOS can resolve repeated iterations that would otherwise require human intervention.

\subsection{Governance Preservation}

AIOS achieved results while maintaining governed execution of the platform.

That includes structured intents, least-privilege Role-Based Access Control (RBAC), governed tool access, human review gates, AIOS review gates where the automation needs clarification, logged actions, proof-of-work artifacts, and reviewable traces and evidence.

The point is that productivity gains did not come from bypassing enterprise controls. The workflow preserved the governance required in regulated financial industries.

\subsection{Summary of Results}

Table \ref{tab:aios-results-by-codebase} summarizes the results by codebase. 

\begin{table}[htbp]
\centering
\caption{Summary of AIOS results by codebase. Pricing is the Pricing Execution Platform. Negotiation is the Client Deal Negotiation Platform.}
\label{tab:aios-results-by-codebase}
\small
\begin{tabular}{lrrr}
\hline
\textbf{Metric} & \makecell[r]{\textbf{Pricing}} & \makecell[r]{\textbf{Negotiation}} & \textbf{Overall} \\
\hline
Enhancement Requests & 30 & 19 & 49 \\
Mean effort reduction & 38\% & 76\% & 53\% \\
Median effort reduction & 40\% & 80\% & 50\% \\
Mean execution cycles & 6.7 & 9.4 & 7.8 \\
Median execution cycles & 5.0 & 4.0 & 5.0 \\
Mean human interventions & 3.0 & 2.6 & 2.9 \\
Median human interventions & 2.0 & 2.0 & 2.0 \\
\hline
\end{tabular}
\end{table}

AIOS produced a measurable reduction in real production workloads. By codebase, the gains were larger in the client deal negotiation platform compared with the pricing execution platform. Human intervention is present in almost every case, supporting the need for humans to steer, review, and provide oversight of AI processes. Governance controls remained intact during request processing.

These results are from real production codebases in the regulated financial industry. They reflect expert-estimated effort reductions, not randomized control trials. The codebases were live and evolving, as was AIOS itself. Interpret the results as descriptive operational evidence of the benefits of AI-assisted software development in regulated industries.

\section{Discussion}\label{sec-discussion}

The results show that AIOS produced meaningful estimated effort reductions in real production workloads. Effort reduction was observed in both codebases:

\begin{itemize}
    \item 38\% mean effort reduction in the pricing execution platform.
    \item 76\% mean effort reduction in the client deal negotiation platform.
    \item 53\% mean effort reduction overall.
\end{itemize}

These findings matter because they were achieved in Zafin's own production environment, which is a regulated financial-services setting rather than in a sandbox, hackathon, or synthetic benchmark.

The most important result is not only that AIOS reduced overall effort, but that it produced these results while maintaining strong governance controls.

Zafin AIOS produced a reduction in effort using real production workloads. Both codebases benefited. The client deal negotiation platform showed greater improvement than the pricing execution platform. There remained human input for reviews and final approval. All requests were performed under governance controls. Zafin AIOS addresses a major challenge for the regulated industry: enabling the use of AI in software development while preserving auditability, control, and human authority.

The 38\% effort reduction in the pricing The 38\% effort reduction in the pricing execution platform, which represents the more complex codebase, should not be interpreted as a shortcoming compared to the client deal negotiation platform. In general, AI can deliver effort reduction also within complex codebases. The negotiation platform suggests that AIOS can add more value in the more repeatable, automatable, or easily specifiable parts of the workflow. That has to be presented as a workload-specific result rather than a general productivity number.

The difference in results between the codebases (38\% and 76\%) is an important finding. It shows that AI-assisted development benefits will vary depending on the codebase, the kinds of tasks, and programming language.

There was human participation for all of the workflows. There were retries and points of human intervention. 3.0 per request in the pricing execution platform. 2.6 per request in the client deal negotiation platform. 2.9 overall. AIOS did not eliminate the human role. Instead, it shifted the human role from direct implementation to intent setting and clarification, steering the AI agents, reviews, and governance oversight.

The average number of execution cycles was 7.8, with 6.7 for the pricing execution platform and 9.4 for the client deal negotiation platform. The client deal negotiation platform had both the highest mean execution cycles and the greatest effort reduction. This suggests that a higher number of interventions count does not necessarily indicate inefficiency. It may reflect AIOS absorbing iterative or stepwise work that would otherwise require additional human effort.

It is important to note that the savings were not through bypassing enterprise controls. This is an important fact because many of the productivity claims are from non-regulated environments. In regulated environments, without controls and auditability, and when dealing with valuable codebases, these AI-assisted tools cannot be used. Zafin AIOS makes Agentic AI useful within regulated environments.

Previous research has identified a gap between experimental AI-assisted development and actual production use in regulated environments. This study provides evidence from live production codebases, handling issues such as stochastic AI behavior, auditability, security, least-privilege access, human-in-the-loop processes, and proof of work. AIOS implements the architectural guidelines identified in the literature review.

We acknowledge multiple constraints and limitations. The baseline relies on expert judgment and team leader consensus. The study does not use randomized controlled trials. The study was not a full parallel, human-led implementation conducted for every task. It was, however, a real implementation with a real production codebase. Both AIOS and the production codebases evolved during the two-month period. The sample includes 49 enhancement requests across two codebases. As an internal study within a single company, generalizability is limited, which is why we describe the results as descriptive. 

The metrics only measure estimated effort reduction. There are additional dimensions such as quality and maintainability that are not captured.

The study delivers valuable insights. A fully controlled parallel implementation would be challenging and likely impractical in a real production codebase. The study's value is found in its applied realism, depicting the complexities intrinsic in actual production environments. Consequently, the results describe operational reality rather than providing definitive proof of causation.


\section{Conclusion}\label{sec-conclusion}

Regulated industries want the speed of AI-assisted software development. However, they cannot adopt unmanaged AI workflows that lack auditability, access control, traceability, and human oversight. The key challenge is moving from AI pilots to governed production use.

This paper presented Zafin AIOS as an orchestration platform and control plane for governed agentic workflows. AIOS combines governed orchestration, proof of work, human authority, granular cost controls, RBAC and controlled access, traceability, and auditability.

The Customer Zero case study examined 49 real enhancement requests. These requests came from two production codebases, the pricing execution platform and the client deal negotiation platform. The study compared AIOS-assisted execution against expert-estimated human-led effort.

AIOS reduced estimated effort in both workloads. Mean effort fell by 38\% for the pricing execution platform, 76\% for the client deal negotiation platform, and 53\% overall. Human intervention remained part of the process, averaging 2.9 interventions per request. Governance controls remained intact throughout the workflows.

The findings suggest that governed AI-assisted software development can produce meaningful effort savings in a regulated production environment such as Zafin. The results support AIOS as a human-supervised augmentation system rather than a fully autonomous replacement for software engineers. The study shows that productivity and governance do not have to be treated as opposing goals.

The contribution is not simply that AI can help write code faster. The contribution is that AI-assisted work can be made observable, auditable, and governable. This is the key requirement for financial services and other regulated industries.

The results are descriptive, not causal proof from a controlled experiment. They are based on expert estimates and team-leader consensus. Interpret the findings as operational evidence from real production workflows. Further studies are needed to validate the results across more teams, codebases, and organizations.

AIOS shows a practical path to move AI-assisted software engineering beyond pilots. By placing AI agents inside governed workflows with proof of work and human authority, enterprises can begin to scale AI-assisted development while preserving the controls required for regulated production software.

\bibliography{refs-used}

\end{document}